# *SEAtech*: Deception Techniques in Social Engineering Attacks: An Analysis of Emerging Trends and Countermeasures

**Devendra Chapagain**
*Department of Computer*
*Birendra Multiple Campus*
*Chitwan, Nepal*
*ORCID: 0009-0003-6548-7112*
devendra.chapagain@bimc.tu.edu.np

**Naresh Kshetri**
*School of Business & Technology*
*Emporia State University*
*Emporia, Kansas, USA*
*ORCID: 0000-0002-3282-7331*
nkshetri@emporia.edu

**Bindu Aryal**
*Department of Computer*
*Birendra Multiple Campus*
*Chitwan, Nepal*
*ORCID: 0009-0006-0216-7500*
bindu.aryal@bimc.tu.edu.np

**Bhawani Dhakal**
*United Technical College, Bharatpur*
*Chitwan, Nepal*
*ORCID: 0000-0002-5375-4290*
dhakal.vawani1@gmail.com

**Abstract**
Social Engineering (SE) is the act of manipulating individuals to perform actions or reveal information. Social engineering tactics are widely recognized as a significant risk to information security. The increasing digital environment has increased the prevalence of social engineering attacks, bringing huge threats to both people and organizations. This paper explores current deception techniques used during social engineering attacks to understand emerging trends and discuss effective countermeasures. It is always a good idea to have knowledge of counter measures and risks from these increasing cyber threats. We have also explored the types of deception attacks and role of social engineering in Advanced Persistent Threats (APTs). Today's major concern for cybersecurity and other web-related attacks is due to social engineering attacks that is also the driving force of increasing cybercrimes worldwide. By uncovering emerging trends and analyzing the psychological underpinnings of these attacks this paper highlights the known deception techniques, emerging trends and counter measures of social engineering attacks.

## 1. Introduction

Social engineering is the psychological manipulation of people into performing actions or exposing confidential information. It is an exploitation of human weaknesses, not technical ones. This is because deception techniques in social engineering have become more sophisticated, through harnessing advancements in technology and human behavioral analysis in crafting highly convincing and targeted attacks. These include phishing and pretexting attacks, moving their way up to more complex methods of attack, like spear-phishing, and deep-fake technology. Alarmingly, these advancements have led to a staggering 65% success rate for social engineering attacks [1]. Even more concerning, social engineering is the catalyst for 93% of successful data breaches [2]. The increased use of social media, global rise of online devices, cybercrime, and other platforms, and the availability of data, further enables the attackers to go about their strategies tailored to specified targets with a lot of detail, making deception very effective [3].

---

\* Correspondence to Devendra Chapagain, *devendra.chapagain@bimc.tu.edu.np*

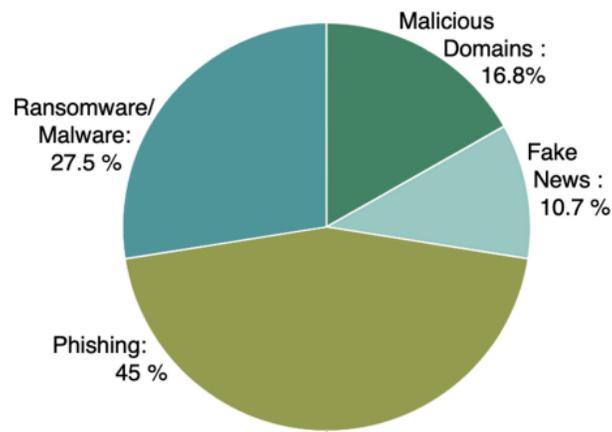

Figure 1: 45 % of COVID-19 Social Engineering Attacks are Phishing Scams (source: [4])

Social engineering attacks aim at deception and fraud for purposes of obtaining sensitive information, money, or computer system access. They involve playing on human emotions and trust to evade security. If any of these red flags are encountered, it is best to exercise caution with also attacks on automobiles and vehicular networks [5]. Avoid clicking on links, opening attachments, or providing personal information. Instead, verify the legitimacy of the request by contacting the supposed sender directly through a trusted channel, such as a phone number or website listed on an official company document.

Deception techniques are a core component of social engineering attacks, designed to exploit the human element, which is often regarded as the weakest link in security systems [6]. Social engineers leverage psychological manipulation to deceive individuals, aiming to influence their beliefs, decisions, and actions in a way that benefits the attacker. This manipulation exploits various psychological vulnerabilities, such as trust, fear, greed, and the desire for social acceptance, to achieve malicious objectives.

Deception serves as a powerful tool for attackers and defenders in cybersecurity. By understanding and mimicking these techniques, defenders can develop effective countermeasures and training programs to improve human resilience. Strategies include simulating phishing attacks, creating honeypots to analyze attacker behavior, and implementing anomaly detection systems. Addressing human vulnerabilities within cybersecurity frameworks is crucial. Recognizing and countering these techniques enhances organizational security posture and protection against evolving threats.

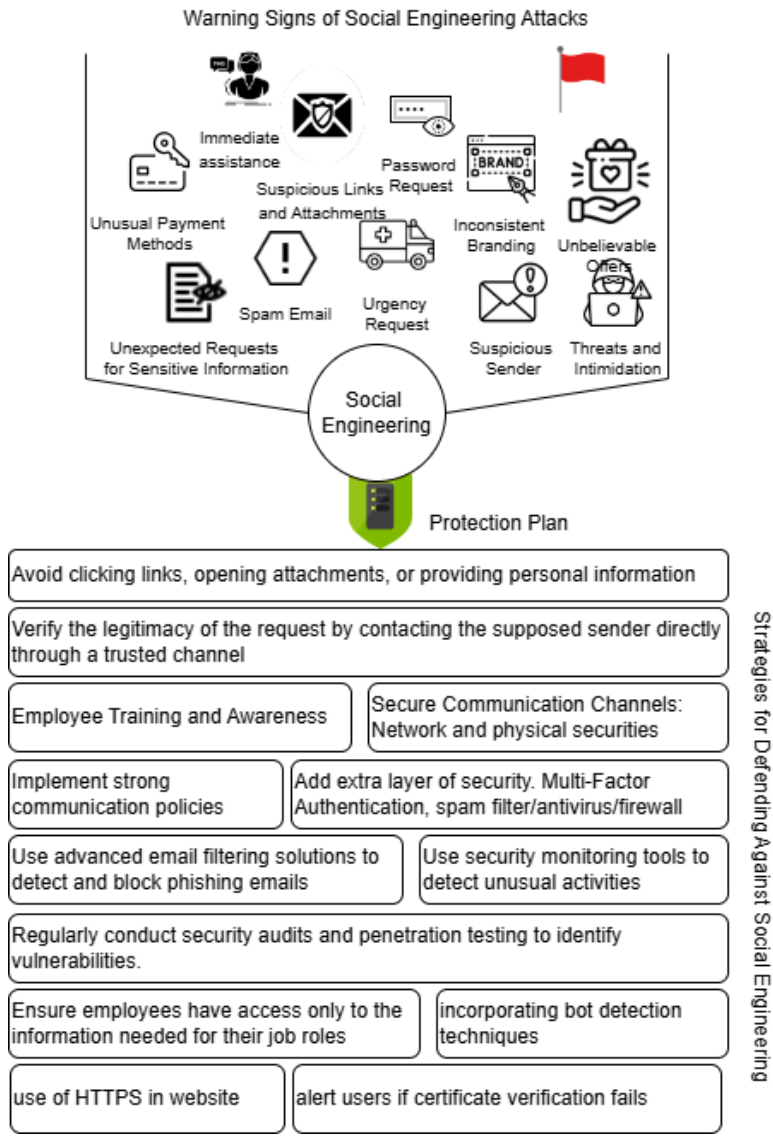

*Figure 2: Taxonomy of social engineering attacks*

Social Engineering attacks are a major cybersecurity concern, representing the dominant force behind cybercrime. Statistics reveal a staggering 98% of cyber-attacks involve some form of SE, exploiting human vulnerabilities rather than technical weaknesses. Unlike malware that targets software flaws, a mere 3% relies solely on technical exploits. Social engineers prey on common human traits like curiosity, trust, politeness, or even apathy, bypassing technical expertise. These attackers, a subset of hackers, specialize in persuasion and influence. They target individuals with access to sensitive information or the power to initiate destructive actions, such as leaking data or launching internal attacks [7].

## 2. Literature Study

A systematic literature review is conducted to identify, analyze, and summarize prevention methods, frameworks, and models for mitigating social engineering attacks. The findings suggest that future research should consider perceived risk and precautionary behavior as predictors of security and privacy on social media platforms [8].

Social engineering is a frighteningly common tactic, with statistics indicating that it was used in a whopping 98% of cyberattacks in 2021 [9]. Recent data analysis has revealed that as high as 66.1% of the total respondents have identified social networking sites as the most prevalent vector with the attack, followed by baiting with 7.34% and phishing with 6.21%. This validates previous studies, bringing to the fore the weakness of social media networks being exploited for malicious intent. Also, 43% of the IT professionals have complained that they had fallen prey to such social engineering attacks in the last year-it shows a massive gap in the defense system and training. The statistics also show an equally disturbing void of awareness on the part of the users. 19.2% of the users are still unaware of the type of attack that they must face, while only 49.2% could confidently say that they recognize a hacked system. This, in turn, underlines the urgency for enhanced training programs and robust security policies most of all for new staff, which prove to be very vulnerable. Moreover, the literature depicts a rising trend in increased social engineering attacks, where attempts have risen by over 500% between the first and second quarter of 2018. A holistic analysis of this kind brings forth the crucial need for continuous research and development in cybersecurity measures in a bid to ensure an all-rounded approach to combating emerging tactics of social engineering attacks [10].

Online social deception prevention mechanisms have been extensively studied, with key algorithms and supervised learning methods specifically designed for network structures. Epidemic models are employed to simulate the spread of false information, while matrix factorization algorithms are used to detect social spammers. Additionally, graph optimization and anomaly detection algorithms are tailored to address the unique challenges posed by network structures. Supervised learning methods play a crucial role, with systems like Truthy and feature sets developed by Ratkiewicz et al. and Kumar et al. [11] extracting network features to detect astroturfing and hoaxes. The performance of these feature sets is evaluated using random forest classifiers, showcasing their effectiveness in identifying deceptive activities online.

Social engineering attacks do not solely target common individuals and employees but also extend to developers who utilize multiple accounts across various platforms such as GitHub, cloud services, repositories, and communication tools like Zoom and Google Docs [12]. Developers are particularly vulnerable as attackers often upload malicious versions of legitimate packages. A notable instance of this involves a phishing campaign specifically designed to steal credentials from PyPI developers, demonstrating the pervasive threat of social engineering across different user groups and platforms.

In [13] authors investigate the psychological underpinnings of social engineering cyberattacks and explore existing countermeasures. Their work emphasizes how cybercriminals exploit human vulnerabilities to orchestrate these attacks. The paper highlights the importance of understanding human behavior in the context of cybersecurity. The paper also discussed the potential of machine learning-based approaches to effectively counter social engineering attacks.

### 3. Proposed Method

The research was done based on a Systematic Literature Review (SLR) methodology for the identification and critical evaluation of relevant studies through the collection of existing literature, including relevant academic journals. Electronic databases used during the research include Google Scholar, IEEE Xplore, and other relevant academic databases. The search strategy focused on academic papers published between 2020 and 2024. This timeframe was chosen because it represents a critical phase in the evolution of social engineering attacks. Papers proposing or analyzing deception techniques used in social engineering attacks were prioritized. This approach distinguishes this study by offering a comprehensive overview of social engineering attacks through the lens of deception techniques.

## 4. Types of Deception Attacks

Social networks have become ubiquitous in modern society, permeating both our professional and personal lives. This widespread use, however, creates a vulnerability. Information about organizational structures, employee relationships, and even personal lives can be exposed not only to the public but also to potential attackers. The statistics underscore this concern: a staggering 80.8% of the world's population actively uses social media, with 37.6% using it for work purposes [14]. Data shows a continuous rise, with 4.95 billion active social media users globally in 2023, reflecting a 7.07% year-on-year increase from 2022 [15]. Alarmingly, social engineering tactics are present in a significant portion of cyberattacks – statistics suggest that 98% of attacks involve some level of social engineering [16].

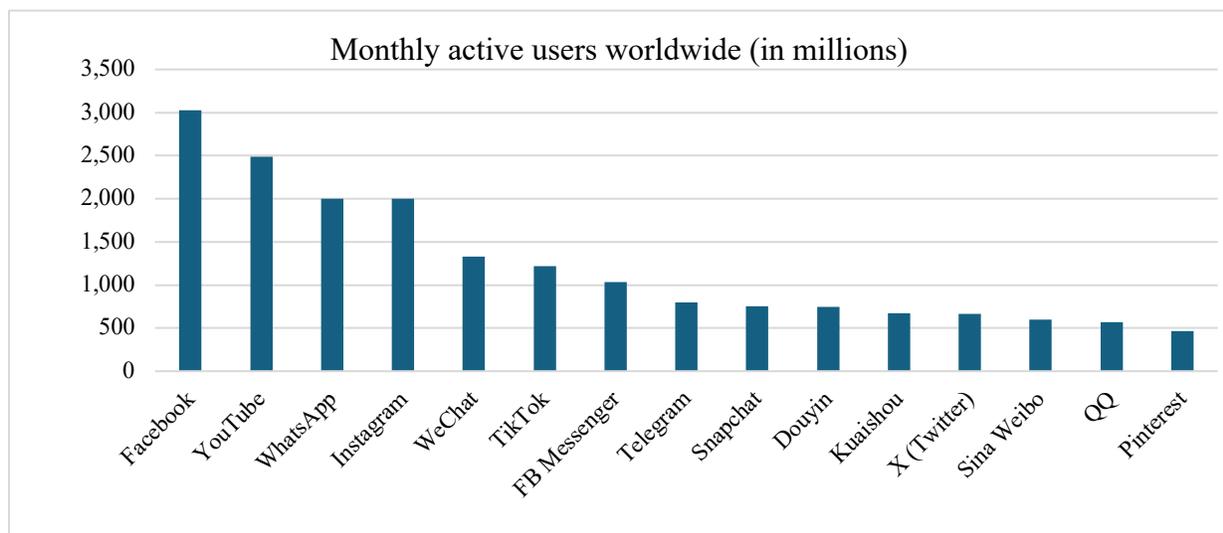

*Figure 3: Monthly Active Users of Leading Social Media Platforms (2024, Source: https://backlinko.com/social-media-users)*

Despite the implementation of various security measures, social engineering attacks continue to rise. This growing threat highlights the importance of understanding the different types of social engineering attacks to effectively combat them.

In [17] authors demonstrated that humans exhibit particularly poor accuracy in detecting deception, with a detection rate of only 18%. Considering the inherent limitations in human judgment, it becomes clear why social engineering is an effective method for bypassing the technical defenses of information security. These findings underscore the vulnerability of relying solely on human intuition to safeguard against social engineering attacks, highlighting the necessity for more robust and comprehensive security measures via several algorithms [18].

**Deception Techniques in Social Engineering (SE) Attacks**

1. Information Manipulation:
   - False information: Fake news, rumors, manipulated information, fake reviews.
   - Luring: Spamming, phishing emails.

2. Identity Theft:
    - Fake identity: Fake profiles, profile cloning, compromised accounts.
    - Crowd-turfing: Creating a false impression of popularity or support.

3. Human-Targeted Attacks: (Exploiting human emotions and trust)
    - Persuasion
    - Impersonation
    - Tailgating or Piggybacking
    - Shoulder Surfing
    - Dumpster Diving

4. Computer-Based Attacks: (Leveraging technology to appear legitimate) [19]
    - Phishing (Often uses email or text messages, with cryptojacking & ransomware attacks)
    - Vishing (Voice phishing - uses phone calls)
    - Watering Hole Attacks (Targets websites frequented by victims)
    - Bot Attacks (Automated attacks using bots)
    - Brand Theft (Imitating a trusted brand to steal information)
    - Baiting (Offering something desirable to lure victims)

## 4.1. The Role of Social Engineering in Advanced Persistent Threats (APTs)

Advanced Persistent Threats (APTs) pose a significant and evolving challenge to cybersecurity [20]. Unlike conventional cyberattacks primarily aimed at immediate financial gain or data destruction, APTs prioritize establishing and maintaining covert access to target networks. These sophisticated attacks often leverage social engineering to bypass technical defenses, exploiting human vulnerabilities to gain an initial foothold. This emphasis on the human element distinguishes APTs from traditional cyber threats, which primarily target system weaknesses. Consequently, organizations must adopt a holistic approach to security, encompassing both technological safeguards and robust employee training programs.

Social engineering serves as the cornerstone of most APT campaigns. These sophisticated attacks often commence with a carefully crafted social engineering attack, designed to infiltrate an organization's network under the radar. By exploiting human vulnerabilities and bypassing traditional security measures, attackers can gain initial access, laying the groundwork for subsequent, more destructive phases of the attack. Spear phishing, a highly targeted form of social engineering, is a prevalent tactic employed by APT groups. These meticulously crafted emails or messages are designed to deceive specific individuals within an organization, often targeting high-level executives or individuals with privileged access. By establishing trust and exploiting curiosity, attackers can induce recipients to click on malicious links, download infected attachments, or reveal sensitive information [21].

Once initial access is gained, attackers leverage social engineering to maintain persistence within the network. They may employ techniques like pretexting, where they impersonate legitimate entities to gain sensitive information or manipulate employees into granting access. By understanding the psychology of their targets and exploiting social norms, APT groups can prolong their presence within a network undetected. While the detection of cyberattacks often relies on publicly available information and social cues, attackers can exploit this same data to craft deceptive campaigns. This adversarial relationship underscores the complexities of modern cybersecurity and necessitates a continuous evolution of defensive strategies.

The sophistication and persistence of APTs highlight the need for a comprehensive security strategy that addresses both technological and human factors. Organizations must invest in advanced detection tools and continuous employee education to effectively counter these threats and protect sensitive information from being compromised.

## 5. Countermeasures for Social Engineering Attacks

The attacks on humans are challenging and more complex to identify. Most of the social engineering attacks succeed and are simply because they are based on the weaknesses of human nature, such as being a trusting person, helping others, and even the fear of missing out. The styles of attacks and strategies most attackers use make it critical to understand the different types of attacks to defend against the attacks in social engineering [22]. Social engineering is protected in so many ways, but the most common and effective are through education, training, and awareness. Figure 2 presented above (Taxonomy of Social Engineering attacks) illustrates the countermeasures of social engineering attacks.

- **Training/education:** To stay ahead of evolving social engineering techniques, organizations should implement ongoing employee education programs on the latest threats and defense mechanisms including knowledge, awareness of computer & cyber. This may include how to keep a strong password and safe online practices [23] [24].
- **Communication policies:** Communication policies serve as a critical line of defense against social engineering attacks by establishing clear guidelines for employee behavior and interaction. By outlining acceptable communication practices, organizations can significantly reduce the risk of falling victim to these attacks.
- **Company equipment:** Company-provided equipment plays a crucial role in bolstering defenses against social engineering attacks. By implementing specific technological measures, organizations can significantly reduce the risk of successful attacks. Firewall and intrusion detection system (IDS), antivirus, email filtering, web filtering and employee monitoring can create a layered defense against social engineering attacks.
- **Spam filters:** Establishing clear guidelines for email usage, such as verifying sender identities, avoiding opening suspicious attachments or clicking on unfamiliar links, and reporting phishing attempts.
- **Encrypted communication:** Encrypted communication provides a robust defense against social engineering attacks by safeguarding sensitive information from unauthorized access and interception. Through the application of cryptographic algorithms, encryption renders data unintelligible to those without the appropriate decryption key, significantly mitigating the risk of successful attacks. The utilization of encryption tools and protocols, such as Virtual Private Networks (VPNs), cloud storage with encryption, and secure email protocols like PGP and S/MIME, is crucial in establishing a secure communication infrastructure [25].

- **Password/data management:** Enforcing the use of strong, unique passwords for each account, coupled with regular password resets, is crucial in mitigating social engineering attacks. By implementing robust password management practices, organizations can significantly reduce the risk of unauthorized access to sensitive information. incident reporting
- **Auditing:** Security audits, vulnerability assessments, penetration testing, and user behavior analytics are essential components of a comprehensive cybersecurity strategy. By conducting these assessments regularly and leveraging the insights gained, organizations can proactively identify and address vulnerabilities, enhancing their resilience against social engineering attacks [26].
- **Cybersecurity policies:** By establishing clear guidelines and expectations, organizations can significantly reduce the risk of employees falling victim to these deceptive tactics.
- **Regular backups:** Implementing regular data backups and disaster recovery plans to protect against data loss along with monitoring & systems security [27].
- **Social media uses:** Providing guidelines for employee social media use, including restrictions on sharing personal and work-related information.

## 6. Conclusion and Future Scope

A systematic literature review was conducted to comprehensively examine the current landscape of social engineering attacks and their corresponding countermeasures. Phishing and malware attacks emerged as the most prevalent forms of social engineering within the studied corpus. While the research identified the potential for diverse social engineering and deception techniques, it primarily focused on the prevalence and nature of these attacks rather than developing a comprehensive framework for countermeasures. To effectively address the evolving threat posed by social engineering, future research should prioritize the development of robust cybersecurity frameworks that incorporate a combination of technical controls, behavioral interventions, and organizational resilience. There are several other cyberattacks (MITM/MITB attacks, cross-site scripting attacks, DDoS attacks, replay attacks, spoofing attacks, cross-site request forgery attacks) that directly occur due to social engineering attacks which can be researched in future.

## References

[1] Team, Marketing. "3 Lessons from the Facebook and Google Loss of $100M to a Spear Phishing Attack | Graphus." Graphus, 21 Jan. 2020, www.graphus.ai/blog/3-lessons-from-the-facebook-and-google-loss-of-100m-to-a-spear-phishing-attack/. Accessed 15 June 2024.

[2] McNeal, Amy. "What Is Social Engineering? Techniques & Prevention | Graphus." Graphus, 22 Apr. 2022, www.graphus.ai/blog/what-is-social-engineering/. Accessed 15 June 2024.

[3] Kshetri, N. (2022). *The global rise of online devices, cyber-crime and cyber defense: Enhancing ethical actions, counter measures, cyber strategy, and approaches*. University of Missouri-Saint Louis. https://irl.umsl.edu/dissertation/1155/

[4] Mashtalyar, N., Ntaganzwa, U.N., Santos, T., Hakak, S., Ray, S. (2021). Social Engineering Attacks: Recent Advances and Challenges. In: Moallem, A. (eds) HCI for Cybersecurity, Privacy and Trust. HCII 2021. Lecture Notes in Computer Science(), vol 12788. Springer, Cham. https://doi.org/10.1007/978-3-030-77392-2_27

[5] Kshetri, N., Sultana, I., Rahman, M. M., & Shah, D. (2024). DefTesPY: Cyber defense model with enhanced data modeling and analysis for Tesla company via Python Language. *arXiv preprint arXiv:2407.14671*.

[6] Zhang, L., & Thing, V. L. L. (2021). Three decades of deception techniques in active cyber defense - Retrospect and outlook. Computers & Security, 106, 102288. doi:10.1016/j.cose.2021.102288

[7] Akyeşilmen, N., & Alhosban, A. (2024). Non-Technical Cyber-Attacks and International Cybersecurity: The Case of Social Engineering. Gaziantep University Journal of Social Sciences, 23(1), 342-360. https://doi.org/10.21547/jss.1346291

[8] Syafitri, W., Shukur, Z., Asma'Mokhtar, U., Sulaiman, R., & Ibrahim, M. A. (2022). Social engineering attacks prevention: A systematic literature review. IEEE access, 10, 39325-39343.

[9] Tulkarm, P. (2021). A survey of social engineering attacks: Detection and prevention tools. *Journal of Theoretical and Applied Information Technology*, *99*(18).

[10] Parsaei, A. (2024). Awareness and Social Engineering-Based Cyberattacks. International Journal of Reliability, Risk and Safety: Theory and Application, 7(1), 31-36.

[11] Guo, Z., Cho, J. H., Chen, R., Sengupta, S., Hong, M., & Mitra, T. (2020). Online social deception and its countermeasures: A survey. Ieee Access, 9, 1770-1806.

[12] "Phishing Campaign Targets PyPI Users to Distribute Malicious Code." Darkreading.com, 2022, www.darkreading.com/cloud-security/phishing-campaign-targets-pypi-users-to-distribute-malicious-code. Accessed 17 June 2024.

[13] Siddiqi, M. A., Pak, W., & Siddiqi, M. A. (2022). A study on the psychology of social engineering-based cyberattacks and existing countermeasures. Applied Sciences, 12(12), 6042.

[14] Wang, Z., Sun, L., & Zhu, H. (2020). Defining social engineering in cybersecurity. *IEEE Access*, *8*, 85094-85115.

[15] "Social Network Usage & Growth Statistics: How Many People Use Social Media in 2024?" Backlinko, Backlinko, 19 Dec. 2023, backlinko.com/social-media-users. Accessed 16 June 2024.

[16] "30 Social Engineering Statistics – 2023." Firewall Times, 18 Sept. 2023, firewalltimes.com/social-engineering-statistics/. Accessed 16 June 2024.

[17] Qin, T., & Burgoon, J. K. (2007, May). An investigation of heuristics of human judgment in detecting deception and potential implications in countering social engineering. In 2007 IEEE Intelligence and Security Informatics (pp. 152-159). IEEE.

[18] Kshetri, N., Kumar, D., Hutson, J., Kaur, N., & Osama, O. F. (2024, April). algoXSSF: Detection and analysis of cross-site request forgery (XSRF) and cross-site scripting (XSS) attacks via Machine learning algorithms. In *2024 12th International Symposium on Digital Forensics and Security (ISDFS)* (pp. 1-8). IEEE.

[19] Kshetri, N., Rahman, M. M., Sayeed, S. A., & Sultana, I. (2024, May). cryptoRAN: A review on cryptojacking and ransomware attacks wrt banking industry-threats, challenges, & problems. In *2024 2nd International Conference on Advancement in Computation & Computer Technologies (InCACCT)* (pp. 523-528). IEEE.

[20] Ö. Aslan, Semih Serkant Aktuğ, Merve Ozkan-Okay, Abdullah Asim Yilmaz, and E. Akin, "A Comprehensive Review of Cyber Security Vulnerabilities, Threats, Attacks, and Solutions," Electronics, vol. 12, no. 6, pp. 1333–1333, Mar. 2023, doi: https://doi.org/10.3390/electronics12061333.

[21] Xing, K., Li, A., Jiang, R., & Jia, Y. (2020). A Review of APT Attack Detection Methods and Defense Strategies. 2020 IEEE Fifth International Conference on Data Science in Cyberspace (DSC). doi:10.1109/dsc50466.2020.00018

[22] Kshetri, N., & Sharma, A. (2021). A review and analysis of online crime in pre and post COVID scenario with respective counter measures and security strategies. *Journal of Engineering, Computing, and Architecture*, *11*(12), 13-33. https://www.researchgate.net/publication/356728307_A_review_and_analysis_of_online_crime_in_pre_and_post_COVID_scenario_with_respective_counter_measures_and_security_strategies

[23] Kshetri, N., & Hoxha, D. (2023, November). knowCC: Knowledge, awareness of computer & cyber ethics between CS/non-CS university students. In *2023 International Conference on Computing, Communication, and Intelligent Systems (ICCCIS)* (pp. 298-304). IEEE.

[24] Zaoui, M., Yousra, B., Yassine, S., Yassine, M., & Karim, O. (2024). A Comprehensive Taxonomy of Social Engineering Attacks and Defense Mechanisms: Towards Effective Mitigation Strategies. IEEE Access.

[25] QBSS. "Safeguarding Your Business: Preventing Social Engineering Attacks - QBSS." QBSS, 20 Oct. 2023, www.quatrrobss.com/articles-blogs/safeguarding-your-business-preventing-social-engineering-attacks/. Accessed 30 July 2024.

[26] "How does Penetration Testing and Vulnerability Assessment improve security? | Intercity.technology," Intercity.technology, 2021. https://www.intercity.technology/solutions/compute-storage/penetration-testing-and-vulnerability-assessment (accessed Jul. 30, 2024).

[27] Kshetri, N., Mishra, R., Rahman, M. M., & Steigner, T. (2024, April). HNMblock: Blockchain technology powered Healthcare Network Model for epidemiological monitoring, medical systems security, and wellness. In *2024 12th International Symposium on Digital Forensics and Security (ISDFS)* (pp. 01-08). IEEE.